\documentclass[journal]{IEEEtran}
\usepackage{times}
\usepackage{amssymb}
\usepackage{helvet}
\usepackage{mathrsfs}
\usepackage{url}
\usepackage{courier}
\usepackage{booktabs,caption}
\usepackage{threeparttable}
\usepackage{amsmath}
\usepackage{float}
\usepackage{graphicx}
\usepackage{booktabs}
\usepackage{url}	
\usepackage{tabularx}
\usepackage{overpic}
\usepackage{color}
\usepackage[english]{babel}
\usepackage[T1]{fontenc}
\usepackage[utf8]{inputenc}
\usepackage{multirow}
\usepackage{diagbox}
\usepackage{authblk}
\usepackage[numbers, compress]{natbib}
\usepackage{times}
\usepackage{epsfig}
\usepackage{graphicx}
\usepackage{amsmath}
\usepackage{amssymb}
\usepackage{caption}
\usepackage{subcaption}
\usepackage[normalem]{ulem}
\usepackage{booktabs}
\usepackage{multirow}
\usepackage{graphics}
\usepackage{mathrsfs}
\usepackage{cuted}
\usepackage{graphbox}
\usepackage{capt-of}

\usepackage[pagebackref=true,breaklinks=true,colorlinks,bookmarks=false]{hyperref}

\begin{document}
\title{Open-Source Skull Reconstruction with MONAI}

\author{Jianning Li, André Ferreira, Behrus Puladi, Victor Alves, Michael Kamp, Moon-Sung Kim, Felix Nensa \\ Jens Kleesiek, Seyed-Ahmad Ahmadi, Jan Egger

\thanks{We acknowledge CAMed (COMET K-Project 871132), FWF enFaced (KLI 678), FWF enFaced 2.0 (KLI 1044) and KITE (Plattform für KI-Translation Essen) from the REACT-EU initiative (\url{https://kite.ikim.nrw/}). Further, we acknowledge NVIDIA for the support.}

\thanks{
J. Li, A. Ferreira, M.-S Kim, J. Kleesiek and J. Egger are with the Institute for Artificial Intelligence in Medicine (IKIM), Essen University Hospital (AöR), Girardetstraße 2, 45131 Essen, Germany.

M. Kamp, M.-S. Kim, J. Kleesiek and J. Egger are with the Cancer Research Center Cologne Essen (CCCE), University Medicine Essen (AöR), Hufelandstraße 55, 45147 Essen, Germany.

J. Kleesiek is with the German Cancer Consortium (DKTK), Partner Site Essen, Hufelandstraße 55, 45147 Essen, Germany.

J. Li, A. Ferreira and J. Egger are with the Computer Algorithms for Medicine Laboratory (Cafe), Graz, Austria.

J. Li and J. Egger are with the Institute of Computer Graphics and Vision (ICG), Graz University of Technology, Inffeldgasse 16c, 8010 Graz, Austria.

B. Puladi is with the Department of Oral and Maxillofacial Surgery, University Hospital RWTH Aachen, Pauwelsstraße 30, 52074 Aachen, Germany and the Institute of Medical Informatics, University Hospital RWTH Aachen, Pauwelsstraße 30, 52074 Aachen, Germany.

M.-S. Kim and F. Nensa are with the Institute of Diagnostic and Interventional Radiology and Neuroradiology, University Hospital Essen (AöR), Essen, Germany.

S.-A. Ahmadi is with the NVIDIA GmbH, Bavaria Towers - Blue Tower, Einsteinstrasse 172, 81677 Munich, Germany.

V. Alves is with the Center Algoritmi, University of Minho, Braga, Portugal.

M. Kamp is with the Institute for Neuroinformatics, Ruhr University Bochum, Germany, and the Data Science and AI Department, Monash University, Melbourne, Australia.

E-mails: Jianning.Li@uk-essen.de; Jan.Egger@uk-essen.de. Corresponding authors: Jianning Li and Jan Egger}}


\maketitle

\textit{\textbf{Abstract-- }}\textbf{\textit{We present a deep learning model based on an autoencoder for the reconstruction of cranial and facial defects using the Medical Open Network for Artificial Intelligence (MONAI) framework, which has been pre-trained on the MUG500+ and SkullFix dataset. The implementation follows the MONAI contribution guidelines, hence, it can be easily tried out and used, and extended by MONAI users. The primary goal of this paper lies in the investigation of open-sourcing codes and pre-trained deep learning models under the MONAI framework. The pre-trained models generated in this work deliever reasonable results on the cranial and facial reconstruction task and provide an ideal starting-point for other researchers interested in further investigating the topic. We released the codes and the pre-trained model at the official MONAI  ‘research contributions' GitHub repository:  \url{https://github.com/Project-MONAI/research-contributions/tree/master/SkullRec}}. This contribution has two novelties: 1. Pre-training an autoencoder on the MUG500+ and SkullFix dataset for cranial and facial reconstruction using MONAI, and open-sourcing the codes and weights for other MONAI users; 2. Demonstrating that existing MONAI tutorials can be easily adapted to new use cases, such as skull (cranial and facial) reconstruction.}

\begin{IEEEkeywords}
Skull reconstruction, Research contribution, MONAI, Open-source, API, PyTorch, Python, Deep learning, Pre-trained model, Cranial implant design, Cranioplasty, Craniotomy, Craniectomy, CT, Bone, Head, Face.
\end{IEEEkeywords}
\IEEEpeerreviewmaketitle


\begin{figure*}[ht]
\centering
\includegraphics[width=1\linewidth]{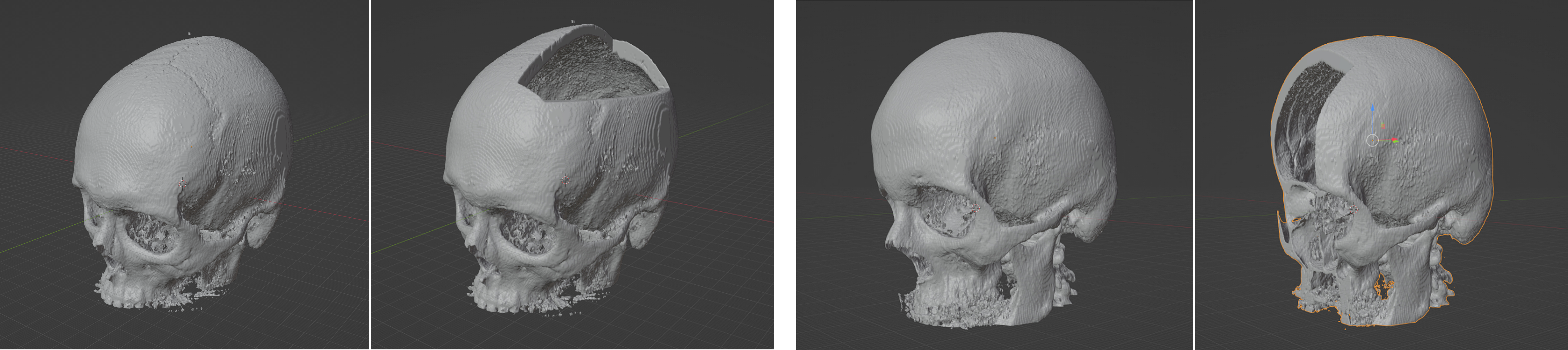}
 \caption{A complete skull (first, third column) and the corresponding skull with a cranial defect (second column) and a facial defect (fourth column) in the MUG500+ dataset.} 
\label{dataset}
\end{figure*}

\section{Introduction}
Nowadays, open-sourcing software, especially (pre-trained) deep learning models, has become increasingly important. Over the years, medical image analysis experienced a tremendous transformation. Over a decade ago, algorithms had to be implemented and optimized with low-level programming languages, like C or C++, to run in a reasonable time on a desktop PC, which was not as powerful as today's computers. Nowadays, users have high-level scripting languages like Python, and frameworks like PyTorch and TensorFlow, along with a sea of public code repositories at hand. As a result, implementations that had thousands of lines of C or C++ code in the past, can now be scripted with a few lines and in addition executed in a fraction of the time. To put this even on a higher level, the Medical Open Network for Artificial Intelligence (MONAI) framework (\url{https://monai.io/} \cite{MonaiWhite}) tailors medical imaging research to an even more convenient process, which can boost and push the whole field. The MONAI framework is a freely available, community-supported, open-source and PyTorch-based framework, that also allows pre-trained models to be integrated as research contributions. To put these claims to the acid test, especially, the point of Easy Integration, we investigate the use of MONAI for open-sourcing pre-trained deep learning models. In particular, we constructed an autoencoder using the MONAI framework, and trained the network for skull reconstruction on the MUG500+ and SkullFix dataset \cite{li2021mug500,kodym2021skullbreak}. The network is trained for reconstructing a complete skull given as input an incomplete skull with cranial or facial defect. We then pushed the codes and the pre-trained autoencoder to the MONAI GitHub repository as a research contribution at \url{https://github.com/Project-MONAI/research-contributions}. 

In our experience, research studies that provide codes and pre-trained models for reproducing the reported results tend to attract much more attention (and therefore citations) than those that do not, benefiting not only the original authors but also other researchers working on the same topic. To our knowledge, deep learning-based craniofacial reconstruction has only recently become known to a small community working on automatic craniofacial implant design, and there still lacks an open-source deep learning library dedicated to craniofacial reconstruction. We believe that integrating such a tool into MONAI, which has already established a decent user base in medical imaging, will broadcast the problem to a wider audience and potentially benefit those willing to further contribute to the topic.

Project MONAI was initiated jointly by NVIDIA and King’s College London, aiming to create an inclusive and community-supported platform, where AI researchers can exchange the best practices of artificial intelligence in healthcare across academia and industry. Among other tools (e.g., MONAI Label), project MONAI created the MONAI framework, which is an open-source and freely available deep learning library that specifically aims at healthcare imaging. It has already been used by researchers, like \cite{godoy2022automatic, belue2022development, shapey2021segmentation}. MONAI provides domain-optimized foundational capabilities for developing healthcare imaging training workflows in a native PyTorch paradigm. In doing so, MONAI features (according to their website):
\begin{itemize}
  \item Open Source Design: MONAI is an open-source project. It is built on top of PyTorch and is released under the Apache 2.0 license.
  \item Standardization: Aiming to capture best practices of AI development for healthcare researchers, with an immediate focus on medical imaging.
  \item User Friendly API: Providing user-comprehensible error messages and easy to program API interfaces.
  \item Reproducibility: Provides reproducibility of research experiments for comparisons against state-of-the-art implementations.
  \item Easy Integration: Designed to be compatible with existing efforts and ease of 3rd party integration for various components.
  \item High Quality: Delivering high-quality software with enterprise-grade development, tutorials for getting started and robust validation \& documentation.
\end{itemize}

Existing MONAI research contributions include, for example, DiNTS (Differentiable Neural Network Topology Search for 3D Medical Image Segmentation) \cite{he2021dints, yu2020c2fnas}, BTCV (3D multi-organ segmentation with UNEt TRansformers) for the Beyond the Cranial Vault challenge) \cite{hatamizadeh2021unetr}, COPLE-Net (COVID-19 Pneumonia Lesion Segmentation) \cite{wang2020noise} and LAMP (Large Deep Nets with Automated Model Parallelism for Image Segmentation) \cite{zhu2020lamp}.

\section*{Code metadata}

\begin{table*}[h!]
\caption{Code metadata}
\centering
\begin{tabularx}{\textwidth}{lX}
\toprule
\textbf{Current code version}   &  1.0 \\
\textbf{Permanent link to code/repository used for this code version} &  \url{https://github.com/Project-MONAI/research-contributions/tree/main/SkullRec}\\
\textbf{Permanent link to Reproducible Capsule}  &  -  \\
\textbf{Legal Code License} & CC-BY-NC-SA  \\
\textbf{Code versioning system used} & none  \\
\textbf{Software code languages, tools, and services used} &  python \\
\textbf{Compilation requirements, operating environments and dependencies} & monai  version: 0.8.1, pytorch version: 1.11.0  \\
\textbf{If available Link to developer documentation/manual} & none  \\
\textbf{Support email for questions} & jianning.li@uk-essen.de  \\
\bottomrule
\end{tabularx}
\label{table:imageInfo}
\end{table*}


\section{Method}
This section describes the main workflow of the contribution, including dataset preprocessing, data format conversion, data resizing and model configurations as well as details about model training, implemented within the MONAI framework.

\subsection{Dataset and Preprocessing}
Two datasets are involved in this study. (1) the MUG500+ dataset \cite{li2021mug500}, which contains 500 complete skulls (e.g., first and third column, Figure \ref{dataset}) in NRRD format. 21 of the image files were found to be corrupted, and were discarded. The remaining 479 complete skulls were split into a training set with 429 skulls and a test set with 50 skulls. Of the 429 complete skulls in the training set, 253 skulls are inserted with cranial defects (e.g., second column, Figure \ref{dataset}) and 176 skulls with facial defects (e.g., fourth column, Figure \ref{dataset}). Correspondingly, we created two defective cases (one with cranial defect and one with facial defect) for each of the 50 test skulls, resulting in $50\times 2=100$ defective test skull samples. The size, shape and position of the defects vary on different skulls. A \textit{complete-defect} skull pair comprises of a defective skull (with either a cranial or a facial defect) and the corresponding complete skull. Therefore, the training set contains 429 such \textit{complete-defect} pairs, and the test set contains $50\times2=100$ \textit{complete-defect} pairs. The cranial and facial defects were created automatically using the defect creation script in the MONAI \textit{SkullRec} repository. 79 \textit{complete-defect} pairs are further split from the training set for validation during training. The axial dimension of all the skull images were cropped to 256. (2) Besides the MUG500+ dataset, we also trained the MONAI network on the SkullFix dataset \cite{kodym2021skullbreak}, which contains 100 \textit{complete-defect} skull pairs for training and 100 pairs for evaluation. In this study, we replaced the original cranial defects with facial defects created the same way as in the MUG500+ dataset, and trained the MONAI network for automatic facial reconstruction.

\subsection{Data Format Conversion}
\begin{figure}
\centering
\includegraphics[width=1\linewidth]{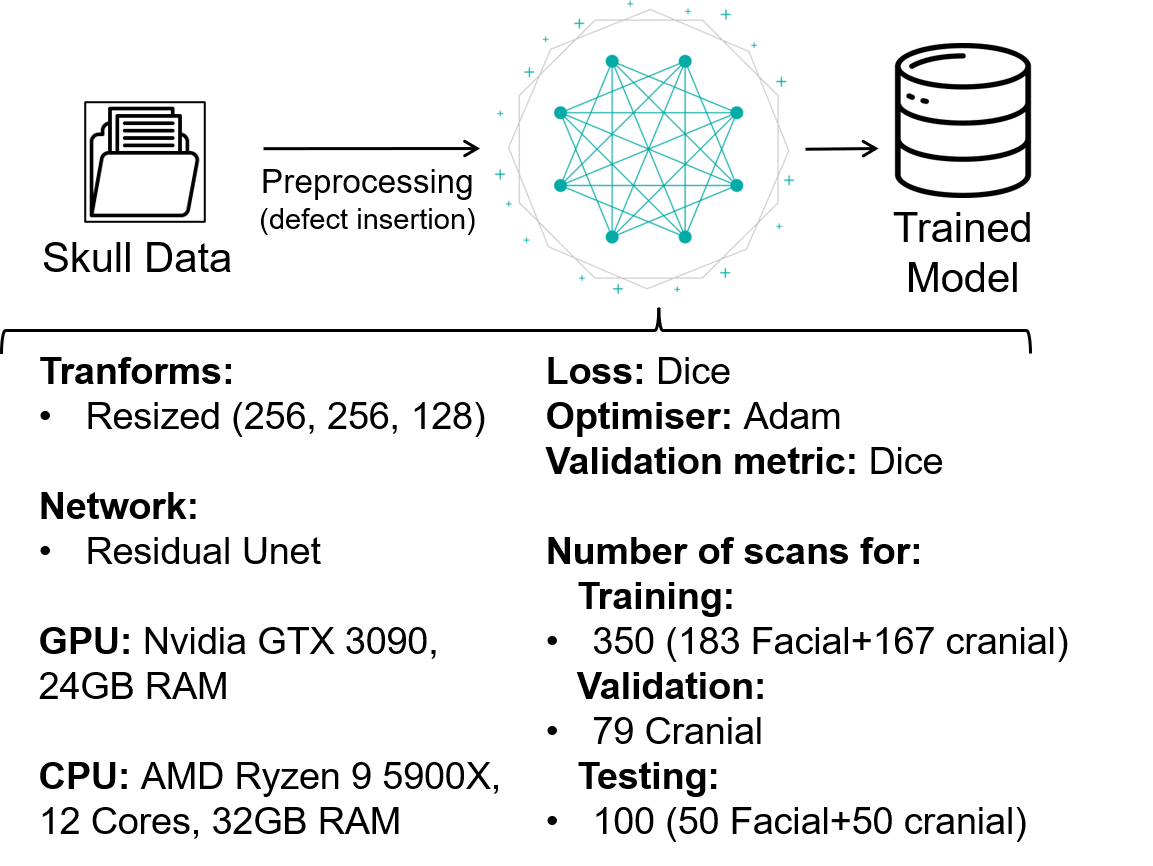}
 \caption{Main workflow for the creation of the trained skull reconstruction model for MONAI, starting with the preprocessing (defect insertion) of the skull dataset.} 
\label{Pipeline1}
\end{figure}

The MONAI framework is optimised for the use of Neuroimaging Informatics Technology Initiative (NIfTI) files due to the integration of the Nibabel \cite{brett_matthew_2020_4295521} library, which makes it easier to use NIfTI files. For this reason, it was necessary to first convert the dataset (which are available in the NRRD file format \cite{Li2022FigshareMUG500Repo}) to the NIfTI format. Note that even though NIfTI is a common format in medical imaging, MONAI's image loader can also directly work with NRRD files. There are several tools that can perform this conversion. We chose to use the Python library Visualization ToolKit (VTK) to perform this conversion \cite{schroeder2003visualization}. Then, the MONAI network is used to train the model on the converted dataset. The final output is the trained model, as shown in Figure \ref{Pipeline1}. 

A download link to the converted and processed dataset is also provided in the MONAI SkullRec Github repository. Users can use the dataset directly for training.

\subsection{Data Resizing and Model Configurations}
\begin{figure*}
\centering
\includegraphics[width=1\linewidth]{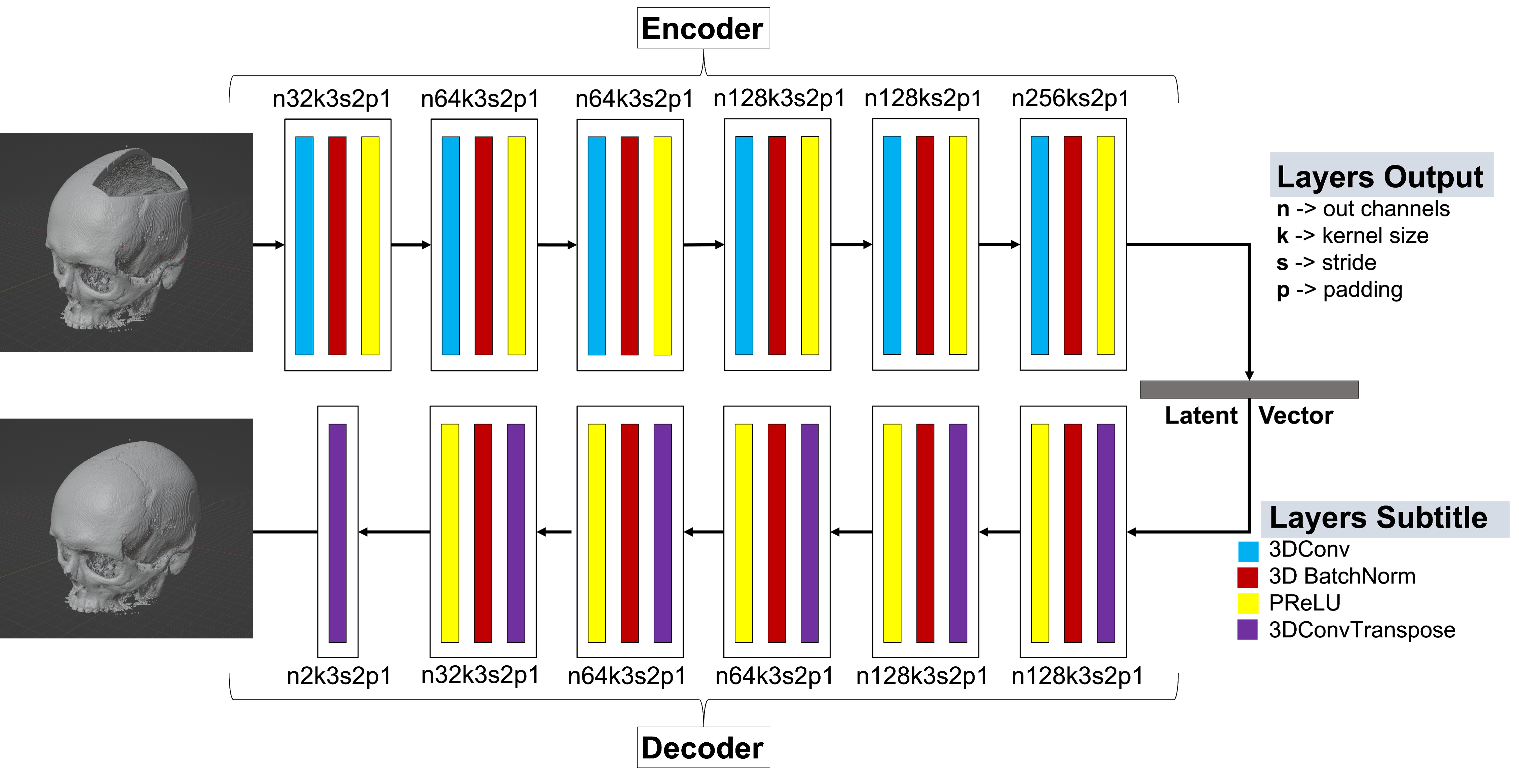}
 \caption{Structure of the autoencoder used for this contribution. The input of the autoencoder is a skull with a synthetic cranial (or facial) defect and the ground truth is the corresponding complete skull.}
\label{AE}
\end{figure*}

The function \textit{Resize} was applied to resize the scans to $256\times256\times128$ using the interpolation mode \textit{Area}. The intensity range scale was not used, because the datasets are already binary. The model was implemented as a simple version of an autoencoder (AE) from the MONAI framework, with the following configurations: spatial\_dims=3; in\_channels=1; out\_channels=2; channels=(32, 64, 64, 128, 128, 256); strides=(2, 2, 2, 2, 2, 2); num\_res\_units=0. The AE is symmetrical, i.e. the downsampling has the same number of layers as the upsampling, as illustrated in Figure \ref{AE}. Major changes to the MONAI framework were avoided by adapting the work pipeline to facilitate further use by others.

\subsection{Model Training}
The autoencoder model described above is trained using the \textit{DiceLoss} loss function and the \textit{Adam} optimiser with a batch size of two,  using respectively the MUG500+ and SkullFix dataset. The function \textit{DiceMetric} is used in the validation step, which is executed every two training epochs. The model is trained for 80 epochs in total, but only the checkpoint corresponding to the best validation score is saved at the end of training. However, it is important to note that, sometimes the validation score can be unreliable, if the validation set is not representative of the dataset. Therefore, it is advised to save not only the best validation checkpoint but also the checkpoint at the last training epoch. When training on the SkullFix dataset, the last checkpoint at the end of training is saved, since SkullFix does not contain a validation set.

To increase the training speed, the MONAI framework provides multithread processing capability to perform data loading and resizing during the training process. It also optimises resource usage by using the GPU for parallel processing, increasing the training speed compared to normal CPU processing. It is important to note that, it is necessary to have the network and data tensors in the same device, in order to avoid data transfer between different devices.

\section{Experiments and Results}

\subsection{Experiments}

We trained the autoencoder specified in Section II (C) using the MUG500+ and SkullFix dataset, respectively, following the model training methods described in Section III (D). On the MUG500+ dataset, the network is trained for reconstructing both cranial and facial defects, since both defect types are included in its training set. On the SkullFix dataset, the network is trained only to reconstruct facial defects. To test the hypothesis that existing MONAI tutorials can be easily adapted to other applications, we directly used the codes provided in a tutorial on 3D spleen segmentation from MONAI.  In this experiment, we only made minor changes to the tutorial codes, including the number of layers in the encoder and decoder and the number of output channels in each layer. The paper aims at conducting a feasibility test on MONAI's open-sourcing and easy-integration feature, instead of competing with state-of-the-art methods on skull reconstruction. Therefore, we provide only a qualitative evaluation of the results, as will be discussed in the following section, by inspecting in 3D the quality of the reconstructed skulls.

\begin{figure*}
\centering
\includegraphics[width=1\linewidth]{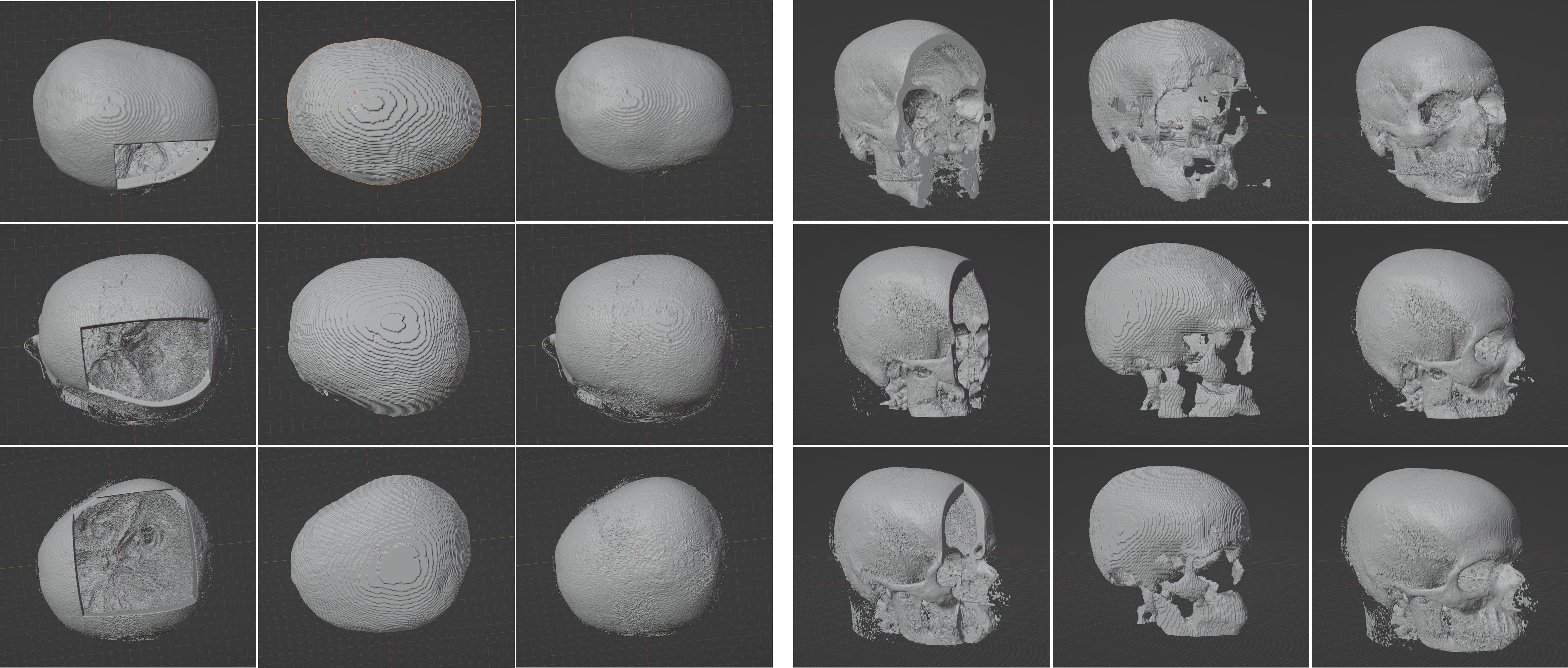}
 \caption{Reconstruction of the cranial (left) and facial (right) defects on the MUG500+ dataset, using the trained model. The first to third column show the input, the prediction and the ground truth, respectively.}
\label{craniofacial_rec}
\end{figure*}

\begin{figure}
\centering
\includegraphics[width=1.0\linewidth]{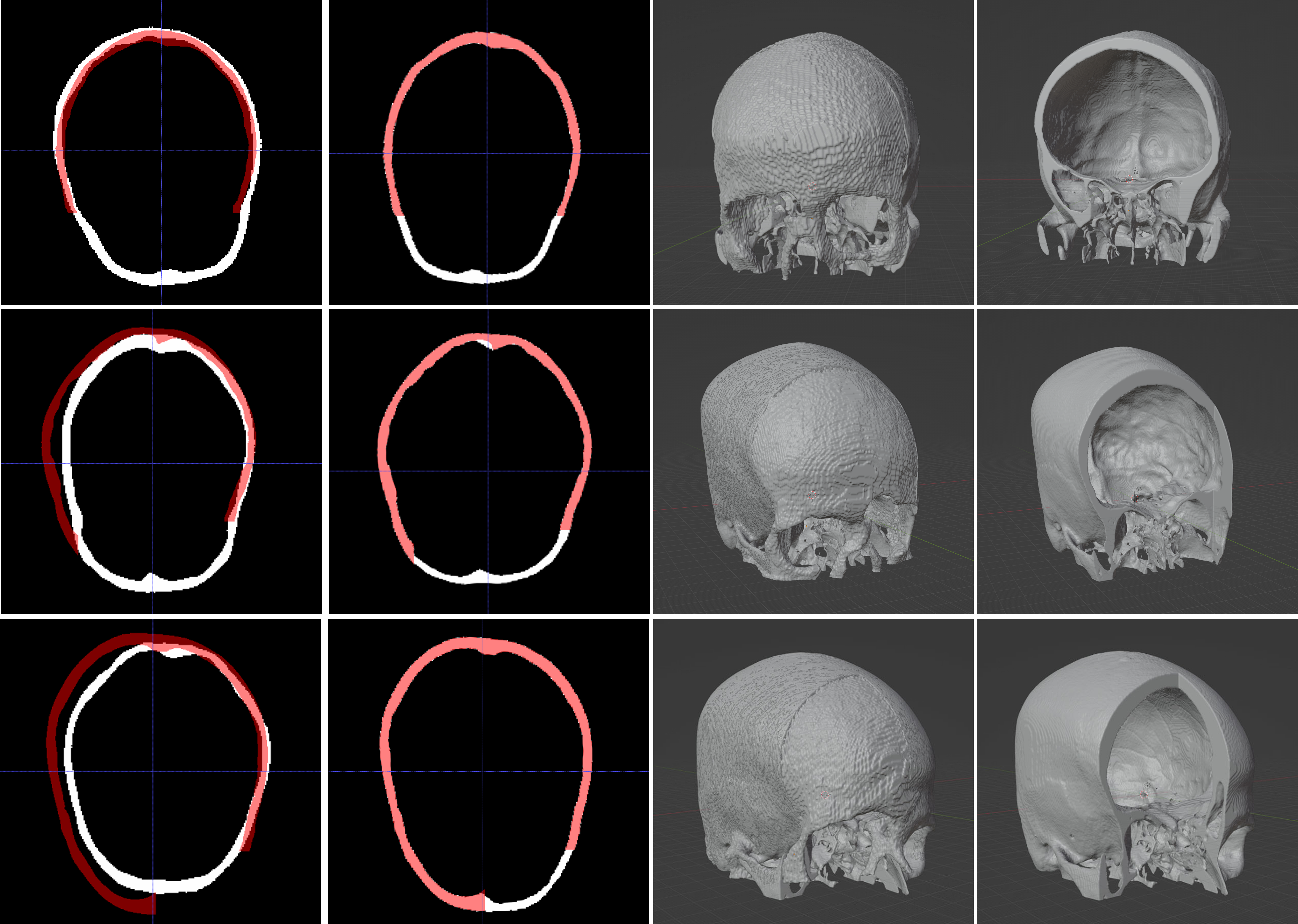}
   \caption{Facial reconstruction results on the SkullFix dataset. The first to the last column shows the axial view of the reconstruction (shown in white) and input (shown in red) before and after alignment, the reconstructed face and input in 3D, respectively.}
\label{monai_results_new}
\end{figure}

\subsection{Results}
Figure \ref{craniofacial_rec} shows the reconstruction of cranial and facial defects for the MUG500+ dataset, using the trained model. We can see that the cranial reconstruction is satisfactory, while the network failed to recover the subtle and complex facial structures. Besides the learning capacity of the network, we attribute the unsatisfactory facial reconstruction performance largely to the MUG500+ dataset itself, as the MUG500+ skulls have obvious artifacts (e.g., spine, catheter extruded from the patients' mouth, etc, as can be seen from the last column of Figure \ref{craniofacial_rec}.) that can potentially distract the network from learning the skull geometries. The removal of these artifacts in a preprocessing procedure is non-trivial, since they are closely connected with the skull, and therefore difficult to be separated from the area of interest (e.g., facial bones). One option to better utilize the dataset for cranial implant design is to crop (axially) and discard the entire facial area of the skulls as in \cite{li2021automatic}. 

On the contrary, the SkullFix dataset contains mostly artifacts-free skulls, and is more suitable for the facial reconstruction task than MUG500+. The last column in Figure \ref{monai_results_new} shows skulls with facial defects. We can see that part or the entire facial structures are missing. The first to third column in Figure \ref{monai_results_new} show the facial reconstruction results obtained using the trained network. Note that the initial reconstruction and the input are misaligned (first column, Figure \ref{monai_results_new}), and therefore the missing facial bones cannot be obtained via a subtraction procedure. We address this by registering the reconstructed completed skull with the input defective skull using a similarity transformation. The second column in Figure \ref{monai_results_new} shows the geometry alignment results. Note that the registration is unsuccessful for some cases. The third column in Figure \ref{monai_results_new} shows the facial reconstruction in 3D. We can see that, compared with the results on MUG500+, the network can learn to restore most of the missing facial structures on the artifacts-free dataset more effectively. Note that the registration step could be avoided by properly aligning the geometry information in the NIfTI output from the model with that of the original NRRD test samples.A 3D viewer for interactive inspection of the skull models can be accessed at \url{https://proj-page.github.io/softwarex_monai.html}.


\section{Discussion}
Deep learning allows to automate tasks that could be done before only manually, or not at all \cite{egger2021deep, EGGER_MedicalMeta}. This also enabled new possibilities in the automatic processing of medical images. As demonstrated in the cranial implant design challenges i.e. AutoImplant I at MICCAI 2020 (\url{https://autoimplant.grand-challenge.org/}) and AutoImplant II at MICCAI 2021 (\url{https://autoimplant2021.grand-challenge.org/}), deep learning-based approaches have shown promising performance in reconstructing skulls where synthetic defects have been inserted. However, deep learning-based approaches still perform unsatisfactorily in reconstructing large and real cranial defects from the clinical routine. In fact, we could recently show that a traditional Statistical Shape Model (SSM) \cite{cootes1992active, heimann2009statistical} can outperform deep learning-based approaches with a fraction of training cases on clinical defective skulls \cite{li2022back}. There are two main reasons for these findings: (1) Synthetic defects only partly resemble real cranial defects, which can be much more complex. In particular, the border area of the defects are more frayed. This, however, can be addressed by creating more realistic-looking synthetic defects. (2) Deep learning-based approaches are data-driven. In general, they work better when more data are involved during the training phase. For the AutoImplant challenges where the SkullFix dataset was used, the participants were provided with only 100 skull pairs for training and we are not aware of participants that used external datasets. The SkullFix dataset is still small for training deep models and a larger skull dataset is desired. A first indication can already be seen from the winning solutions of both AutoImplant challenges \cite{ellis2020deep,wodzinski2021improving}, which drastically expanded the training set through data augmentation. Thus, we assume that our new collection of the MUG500+ skull dataset collection, which was released after the AutoImplant challenges \cite{kodym2021skullbreak, li2021synthetic}, will be an impetus for future advancements of learning-based skull reconstruction methods. However, as discussed in Section III (B), the MUG500+ dataset does not come without its limitations. The facial artifacts that cannot be removed easily pose major challenges for deep learning models to learn the facial geometries effectively.  Nevertheless, since this is the first time the MUG500+ dataset is used for skull reconstruction after its release, the data preprocessing strategies introduced in this paper could server as a sound starting-point for other researchers working on the problem. 

In addition, MONAI allows us to provide and share an easily accessible pre-trained model on the dataset for the community. This can help to disseminate and address remaining challenges in patient-specific and fully-automatic craniofacial implant design. An example is the usage of automatically designed implants in cranioplasty procedures without major modifications \cite{ellis2021qualitative}. Another aspect that has to be addressed by the research community is the implant thickness, which should be thinner than the skull bone. Furthermore, because implants design regulations and techniques can vary among different clinical institutions and countries \cite{campe2020patient, rauschenbach2021personalized}, federated learning could be used to train a model using datasets from various sources.


\section{Conclusion}
In this contribution, we presented a pre-trained autoencoder for automatic skull reconstruction as a MONAI research contribution. We showed that existing open-source MONAI tutorials (e.g., '3D spleen segmentation') can be easily adapted to new applications, such as cranial and facial reconstruction. Reasonable results can be obtained without major modification to the sources. The codes and pre-trained models can also be conveniently shared on the MONAI GitHub platform for other MONAI users interested in this topic. The data sets for the pre-training originate from the MUG500+ dataset \cite{li2021mug500}. All skulls are complete (healthy) with no holes or fractures. Hence, the skulls can be used, for example, by injecting synthetic holes into the healthy skulls and steering an algorithm to learn the task of skull completion \cite{morais2019automated}. A pure end-user and browser-based solution can be tried out in the online framework StudierFenster (\url{www.studierfenster.at} \cite{egger2022studierfenster}) within the 3D Skull Reconstruction module \cite{li2021online}.

Future work sees the pre-training on more data, especially from different clinical institutes that cover a wider variety of CT scanners, scanning protocols, resolutions, etc. Furthermore, a federated learning approach to train algorithms across multiple decentralized edge devices or servers holding local data samples, is desirable, so that researchers can incorporate their own datasets in the pre-trained model, without sharing the dataset. Finally,  A model capable of processing multimodal inputs (texts, images, etc) \cite{heiliger2022beyond}, similar to a real physician, is more desirable for the skull reconstruction task, compared to training on skull images alone. Refer to \cite{li2022training} for another open-source MONAI-based skull reconstruction project about latent space disentanglement \cite{fragemann2022review}.

\section*{Acknowledgement}
We acknowledge CAMed (COMET K-Project 871132), enFaced (FWF KLI 678), enFaced 2.0 (FWF KLI 1044) and KITE (Plattform für KI-Translation Essen) from the REACT-EU initiative (EFRE-0801977, \url{https://kite.ikim.nrw/}).

\bibliographystyle{IEEEtran}
\bibliography{references}
\end{document}